\documentclass[twocolumn,showpacs,superscriptaddress,pre]{revtex4}
\usepackage{graphicx,amssymb,amsmath}

\begin{document}

\title{Dynamo Transition in Low-dimensional Models}

\author{Mahendra K. Verma}

\affiliation{Department of Physics, IIT Kanpur, India}

\author{Thomas Lessinnes}

\affiliation{Physique Statistique et Plasmas, Universit\'e Libre de Bruxelles, B-1050
Bruxelles, Belgium}

\author{Daniele Carati}

\affiliation{Physique Statistique et Plasmas, Universit\'e Libre de Bruxelles, B-1050
Bruxelles, Belgium}

\author{Ioannis Sarris }

\affiliation{Department of Mechanical and Industrial Engineering, University of
Thessaly, Volos, Greece}

\author{Krishna Kumar}

\affiliation{Department of Physics, IIT Kharagpur, India}

\author{Meenakshi Singh}

\affiliation{Department of Physics, Penn-state University, University Park, USA.}

\begin{abstract}
Two low-dimensional magnetohydrodynamic models containing three velocity and three magnetic modes are described. One of them (nonhelical model) has zero kinetic and current helicity, while the other model (helical) has nonzero kinetic and current helicity.
The velocity modes are forced in both these models. 
These low-dimensional models exhibit a dynamo transition at a critical forcing amplitude that depends on the Prandtl number.   In the nonhelical model, dynamo exists only for magnetic Prandtl number beyond 1, while the helical model exhibits dynamo for all magnetic Prandtl number.   Although the model is far from reproducing all the possible features of dynamo mechanisms, its simplicity allows a very detailed study and the observed dynamo transition is shown to bear similarities with recent numerical and experimental results.
\end{abstract}


\maketitle

\section{Introduction}
The understanding of magnetic field generation, usually referred to as the dynamo effect, in planets, stars, galaxies, and other astrophysical objects remains one of the major challenges in turbulence research. There are many observational results from the studies of the Sun, the Earth, and the galaxies~\cite{Moff:book,Krau:book,Bran:PR}. Dynamo has also been observed recently in laboratory experiments~\cite{Fauv:VKS_PRL06,Fauv:dynamo_reverse_EPL06} that have made the whole field very exciting. Numerical simulations~\cite{Sche:dynamo_critPm_PRL04,Sche:dynamo_lowPm_PRL07,Pont:dynamo_lowPr_PRL05,Mini:DynamoPoP06,Mini:LowPr} also give access to many useful insights into the physics of dynamo. However, the complete understanding of the dynamo mechanisms has not yet emerged.

The two most important nondimensional parameters for the dynamo studies are the Reynolds number $Re=UL/\nu$ and the magnetic Prandtl number $P_m=\nu/\eta$, where $U$ and $L$ are the large-scale velocity and the large length-scale of the system respectively, and $\nu$ and $\eta$ are the kinematic viscosity and the magnetic diffusivity of the fluid. Another nondimensional parameter used in this field is the magnetic Reynolds number $Re_m$, defined as $UL/\eta$. Clearly $Re_m=Re\ P_m$, hence only two among the above three parameters are independent. Note that galaxies, clusters, and the interstellar medium have large $P_m$, while stars, planets, and liquid sodium and mercury (fluids used in laboratory experiments) have small $P_m$ \cite{Moff:book,Krau:book}. 

In a typical simulation, the magnetofluid is forced and the dynamo transition is considered to be observed when a nonzero magnetic field is sustained in the steady-state laminar solution or in the statistically stationary turbulent solution, depending on the regime. Typically, dynamos occur for forcing amplitudes beyond a critical value which also defines the critical Reynolds number $Re^{c}$ and the critical magnetic Reynolds number $Re_m^{c}$. One of the objectives of both the numerical simulations and the experiments~\cite{Fauv:VKS_PRL06,Fauv:dynamo_reverse_EPL06} is the determination of this critical magnetic Reynolds number $Re_m^{c}$. It has been found that $Re_m^{c}$ depends on both the type of forcing and the Prandtl number (or Reynolds number), yet the range of $Re_m^{c}$ observed in the numerical simulations is from 10 to 500 for a wide range of $P_m$ (from $5\times10^{-3}$ to 2500).

In recent magnetohydrodynamics (MHD) simulations, Schekochihin et al.~\cite{Sche:dynamo_critPm_PRL04,Sche:DynamoApJ04} applied nonhelical forcings and observed that the dynamo is active for a magnetic Prandtl number larger than a critical Prandtl number $P_m^{c}$ that is around 1. For fluids with small Prandtl number $P_m$ ($P_m<1)$, numerical simulations~\cite{Sche:dynamo_lowPm_PRL07,Pont:dynamo_lowPr_PRL05,Mini:DynamoPoP06} indicate that the dynamo can be produced using forcings having local helicity (the net helicity of the force could still be zero). The range of the critical magnetic Reynolds number in most of the simulations~\cite{Sche:dynamo_lowPm_PRL07,Pont:dynamo_lowPr_PRL05,Mini:DynamoPoP06} is 10 to 500. Note that in the Von-Karman-Sodium (VKS) experiment, the critical magnetic Reynolds number is around 30. 

There are many attempts to understand the above observations. For large Prandtl number, the resistive length scale is smaller than the viscous scale. For this regime, Schekochihin et al.~\cite{Sche:dynamo_critPm_PRL04,Sche:DynamoApJ04} suggested that the growth rate of the magnetic field is higher in the small scales because stretching is faster at these scales. This kind
of magnetic field excitation is referred to as \emph{small-scale turbulent dynamo}. For low $P_m$ Stepanov and Plunian \cite{Step:DynamoGrowth} argue for similar growth mechanism.  Their results are based on shell model calculations.  The numerical results of Iskakov et al.~\cite{Sche:dynamo_lowPm_PRL07}
however are not conclusive in this regard.  The main arguments supporting these explanations are  based on the inertial range (small-scales) properties of turbulence~\cite{Sche:dynamo_critPm_PRL04,Sche:DynamoApJ04}. In this paper, we present an alternate viewpoint. We show that a low-dimensional dynamical system containing only large scales properties of the fields, three velocity and three magnetic Fourier modes, is able to reproduce some of the above numerical results. For  certain types of  forcing of velocity, a dynamo transition is observed for $Re_m>Re_m^{c}$. These observations indicate that the large-scale eddies may also be responsible for the dynamo excitation, and several important properties can be derived from the dynamics of large-scale modes. These observations are consistent with earlier results of MHD turbulence indicating that the large-scale velocity field provides a significant fraction of the energy (around 40\%) contained in the large-scale magnetic field \cite{Dar:flux,Oliv:Simulation,MKV:PR,Mini:I,Mini:II,Olvi:JOT}. 

\section{Derivation of low dimensional models}

The incompressible MHD equations are:
\begin{align}
\partial_t\mathbf{u}&=-\mathbf{n}(\mathbf{u}, \mathbf{u})+\mathbf{n}(\mathbf{b},\mathbf{b})+\nu\nabla^{2}\mathbf{u}+\mathbf{f}-\nabla p_{tot},\label{eq:mhd1}\\
\partial_t\mathbf{b}&=-\mathbf{n}(\mathbf{u}, \mathbf{b})+\mathbf{n}(\mathbf{b}, \mathbf{u})+\kappa\nabla^{2}\mathbf{b},\label{eq:mhd2}\\
\nabla\cdot\mathbf{u}&=\nabla\cdot\mathbf{b}=0,
\label{eq:mhd3}
\end{align}
where $p_{tot}$ is the sum of the hydrodynamic and magnetic pressures divided by the density. The bilinear operator is defined as $\mathbf{n}(\mathbf{a}_1, \mathbf{a}_2)=\mathbf{a}_1\cdot\nabla\mathbf{a}_2$. Only periodic solution in a cubic box with linear dimension $\ell$ will be considered. The smallest non-zero wave vector is thus given by $k_0=2\pi/\ell$. The model is derived by projecting the MHD equations on the subspace ${\cal S}^<$ spanned by a small number of basis vectors that are compatible with the periodic boundary conditions and can be regarded as a subset of a complete basis. Only three vectors will be considered in this study, so that the projection of both the equations for $\mathbf{u}$ and $\mathbf{b}$ on these vectors leads to a six-dimensional system of equations. The three vectors are defined as follows:
\begin{align}
\mathbf{e}_{1}&=\frac{2}{\sqrt{2+h^2}}\ \left(\begin{array}{c} -\sin{k_0x}\ \cos k_0z\\ h\ \sin{k_0x}\ \sin k_0z\\ \cos k_0x\ \sin k_0z\end{array}\right),\\
\mathbf{e}_{2}&=\frac{2}{\sqrt{2+h^2}}\ \left(\begin{array}{c} h \sin k_0y\sin k_0z\\ -\sin k_0y\ \cos k_0z\\ \cos k_0y\ \sin k_0z\end{array}\right),\\
\mathbf{e}_{3}&=\frac{4}{\sqrt{6+10h^2}}\ \left[ \left(\begin{array}{c} -\sin k_0x\ \cos k_0y\ \cos2k_0z\\ -\cos k_0x\ \sin k_0y\ \cos2k_0z\\ \cos k_0x\ \cos k_0y\ \sin2k_0z\end{array}\right)\right.\nonumber\\
&\left. +h\ \left(\begin{array}{c} 0\\ -2 \sin k_0x\ \cos k_0y\ \sin2k_0z\\ \sin k_0x\ \sin k_0y\ \cos 2k_0z\end{array}\right)\right]
\label{eq:basis}
\end{align}
They depend on a free parameter $h$ that determines the amount of helicity that can be carried on by the basis vectors. For $h=0$, each mode is nonhelical. The three vectors are also divergence free $\nabla \cdot \mathbf{e}_\alpha=0$ and are orthonormal $\langle e_\alpha \cdot e_\beta\rangle=\delta_{\alpha \beta}$. The inner product of two arbitrary vectors $\mathbf{v}_1$ and $\mathbf{v}_2$ is defined by:
\begin{equation}
\langle \mathbf{v}_1 \cdot \mathbf{v}_2 \rangle = \frac{1}{\ell^3}\int_0^\ell\!\!\! dx\!\! \int_0^\ell\!\!\! dy\!\!\int_0^\ell\!\!\! dz
\, \mathbf{v}_1(x,y,z)\cdot \mathbf{v}_2(x,y,z)\,.
\end{equation}
For $h=0$, the basis functions $\mathbf{e}_{1}$ and $\mathbf{e}_{2}$ have field configurations in $xz$ and $yz$ respectively.

The projection operator is noted $P$ and its application on a vector $\mathbf{v}$ is defined by:
\begin{equation}
P[\mathbf{v}]\equiv\mathbf{v}^<=v_1\ \mathbf{e}_{1}+v_2\ \mathbf{e}_{2}+v_3\ \mathbf{e}_{3}
\end{equation}
where $v_\alpha=\langle\mathbf{v}\cdot\mathbf{e}_\alpha\rangle$. The projected part of $\mathbf{v}$ in ${\cal S}^<$ is noted $\mathbf{v}^<$ and the difference with the original vector is noted $\mathbf{v}^>=\mathbf{v}-\mathbf{v}^<$. The projection of the velocity, magnetic, and force fields are expressed as follows.
\begin{align}
P[\mathbf{u}]&=\mathbf{u}^< = (u_1\ \mathbf{e}_{1}+u_2\ \mathbf{e}_{2}+u_3\ \mathbf{e}_{3}) \ u^\star,\label{eq:uexpand}\\
P[\mathbf{b}]&=\mathbf{b}^< = (b_1\ \mathbf{e}_{1}+b_2\ \mathbf{e}_{2}+b_3\ \mathbf{e}_{3}) \ u^\star,\label{eq:bexpand}\\
P[\mathbf{f}]&=\mathbf{f}^< = (f_1\ \mathbf{e}_{1}+f_2\ \mathbf{e}_{2}+f_3\ \mathbf{e}_{3})\ (u^\star)^2 \ k_0.\label{eq:fexpand}
\end{align}
where the $u_\alpha$'s, $b_\alpha$'s and $f_\alpha$'s are dimensionless real numbers and $u^\star=\nu k_0$. The $\mathbf{e}_3$ component of the forcing is supposed to be zero. The system of dynamical equations for these variables can be derived from Eqs. (\ref{eq:mhd1}-\ref{eq:mhd3}). For instance the projection of the equation for the velocity on the vector $\mathbf{e}_1$, $\langle \partial_t \mathbf{u}\cdot \mathbf{e}_1\rangle$, yields the equation for $\dot{u_1}$. The resulting system of equations is:
\begin{eqnarray}
\dot{u_1}& = & r(h)\, (1-h^2)\, (u_2u_3-b_2b_3)-2\, u_1+f_1\label{eq:w101}\\
\dot{u_2}& = & r(h)\, (1+3h^2)\,  (u_1u_3-b_1b_3)-2\, u_2+f_2\label{eq:w011}\\
\dot{u_3}& = & -2\, r(h)\, (1+h^2)\, (u_1u_2-b_1b_2)-6\, u_3+f_3\label{eq:w112}\\
\dot{b_1}& = & r(h)\, (1+h^2)\, (u_2b_3-b_2u_3)-2\, P_m^{-1}\, b_1\label{eq:b101}\\
\dot{b_2}& = & -r(h)\, (1+h^2)\, (u_3b_1-b_3u_1)-2\, P_m^{-1}\, b_2\label{eq:b011}\\
\dot{b_3}& = & -2\, r(h)\, h^2\, (u_1b_2-b_1u_2)-6\, P_m^{-1}\, b_3.\label{eq:b112}
\end{eqnarray}
where $r(h)=2/\left((2+h^2)\ \sqrt{6+10h^2}\right)$. 

The structure of this model is, of course, reminiscent of the original MHD equations. In particular, it is interesting to study how the conservation of the ideal quadratic invariants of the Navier-Stokes and MHD equations can be expressed in terms of the variables \{$u_\alpha$, $b_\alpha$\}. The conservation of the kinetic energy $E_k$ by the Navier-Stokes equations is a direct consequence of 
\begin{equation}
\langle \mathbf{n}(\mathbf{u}, \mathbf{u}) \cdot \mathbf{u} \rangle =0
\end{equation}
Because this equality holds for any velocity, it is also true for $\mathbf{u}^<$ and, in the absence of magnetic field, the nonlinear terms in the system of equations~(\ref{eq:w101}-\ref{eq:b112}) conserve the kinetic energy $E_k^<=(u_1^2+u_2^2+u_3^2)/2$ associated to $\mathbf{u}^<$. The cross helicity $H_c^<=(u_1b_1+u_2b_2 +u_3b_3)$ and the total energy, the sum of the kinetic energy $E_k^<$ and the magnetic energy $E_m^<=(b_1^2+b_2^2+b_3^2)/2$, are conserved by the nonlinear terms for the same reason. However, the conservation of the kinetic helicity $H_k=\langle \mathbf{u}\cdot \mathbf{\omega} \rangle$ by the Navier-Stokes equations and the conservation the magnetic helicity $H_m=\langle \mathbf{b}\cdot\mathbf{a} \rangle$ by the MHD equations, where $\mathbf{\omega}=\nabla\times\mathbf{u}$ is the vorticity and $\mathbf{a}$ is the vector potential ($\mathbf{b}=-\nabla\times\mathbf{a}$) have no equivalent in the system~(\ref{eq:w101}-\ref{eq:b112}). Indeed, the conservation of the kinetic helicity is a consequence of 
\begin{equation}
\langle \mathbf{n}(\mathbf{u},\mathbf{u}) \cdot \mathbf{\omega} \rangle =0
\end{equation}
However, both the nonlinear term and the vorticity are not fully captured in ${\cal S}^<$, even when computed from $\mathbf{u}^<$: 
\begin{align}
P[\mathbf{n}(\mathbf{u}^<, \mathbf{u}^<)]&\not=\mathbf{n}(\mathbf{u}^<, \mathbf{u}^<)\,,\\
P[\nabla\times\mathbf{u}^<]&\not=\nabla\times\mathbf{u}^<\,,
\end{align}
As a consequence, the kinetic helicity carried on by $\mathbf{u}^<$ is not conserved by the nonlinear term in the low-dimensional model and, for the same reason, the magnetic helicity is not conserved either. For instance, the non-conservation of the kinetic helicity by the low-dimensional model can be expressed as,
\begin{equation}
\langle \mathbf{n}(\mathbf{u}^<,\mathbf{u}^<)^< \cdot (\nabla\times\mathbf{u}^<)^< \rangle = - \langle \mathbf{n}(\mathbf{u}^<,\mathbf{u}^<)^> \cdot (\nabla\times\mathbf{u}^<)^> \rangle\,, 
\end{equation}
and the right hand side in this relation in not represented in the equations~(\ref{eq:w101}-\ref{eq:b112}).
In the following section we will solve the truncated MHD equations ~(\ref{eq:w101}-\ref{eq:b112}) under special cases.

\section{Analysis of the model} \label{Sec:Analysis}

Despite its simplicity, the complete analysis of the system of equations~(\ref{eq:w101}-\ref{eq:b112}) is quite difficult. Indeed, the simple search for the fixed points leads to very intricate algebraic equations. We have thus focused our analysis on two limit cases corresponding to strictly nonhelical equations ($h=0$) and the case $h=1$,  both of which can be treated analytically.

\subsection{Nonhelical model}

The system of equations~(\ref{eq:w101}-\ref{eq:b112}) is thus first considered for $h=0$. In that case the kinetic helicity captured in subspace ${\cal S}^<$ vanishes ($\langle \mathbf{u}^< \cdot (\nabla \times \mathbf{u}^<)\rangle=0$). In this first model, the forcing is chosen to be
\begin{equation}
\mathbf{f}=\frac{f}{\sqrt{2}}\, (\mathbf{e}_1+\mathbf{e}_2)
\end{equation}
where $f=\sqrt{\langle \mathbf{f}\cdot\mathbf{f}\rangle}$. Hence, $f_1=f_2=f/\sqrt{2}$ and $f_3=0$. The fixed points can then be computed analytically. Three fixed points correspond to a vanishing magnetic ($b_1=b_2=b_3=0$) field and will be referred to as fluid solutions:
\begin{align}
\emph{Fluid A$^\pm$}&
\left\{
\begin{array}{l}
u_1=\displaystyle{\frac{\sqrt{2}}{8}}\ \left(f\pm\sqrt{f^2-1152}\right)\\
\\
u_2=\displaystyle{\frac{\sqrt{2}}{8}}\ \left(f\mp\sqrt{f^2-1152}\right)\\
\\
u_3=-2\sqrt{6}
\end{array}
\right.
\label{eq:SolnA}\\
\nonumber\\
\emph{Fluid B}&
\left\{
\begin{array}{l}
u_1=u_2=\displaystyle{\frac{(s^2-12)}{s}}\\
\\
u_3=-\displaystyle{\frac{(s^2-12)^2}{\sqrt{54}\ s^2}}
\end{array}
\right.
\label{eq:solnB}
\end{align}
where $s=\left(9f/\sqrt{2}+3\sqrt{192+9f^{2}/2}\right)^{1/3}$. There are two additional fixed points with nonzero magnetic fields:
\begin{equation}
\label{eq:SolnC}
\emph{MHD$^\pm$}
\left\{
\begin{array}{l}
u_1=u_2=\displaystyle{\frac{\sqrt{2}f P_m}{4(P_m+1)}}\\
\\
u_3 = -\displaystyle{\frac{2\sqrt{6}}{P_m}}\\
\\
b_1=b_2=\displaystyle{\pm\frac{P_m}{4(P_m+1)}\ \sqrt{f^2-f_{c_2}^2}}\\
\\
b_3=0
\end{array}\right.
\end{equation}

\begin{figure}[ht]
\begin{centering}
\includegraphics[width=0.8\columnwidth,keepaspectratio]{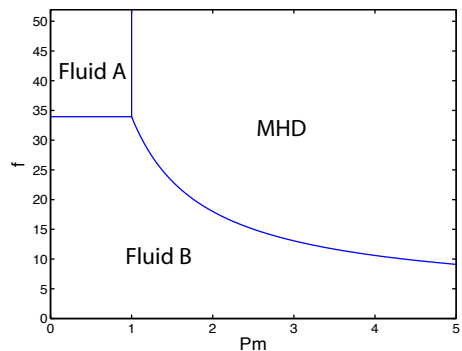}
\end{centering}
\caption{Plot of critical force $f_{c_2}$ as a function
of $P_m$ for the nonhelical model.  Dynamo is excited for $f>f_{c_2}$, only for $P_m>1$.}.
\label{fig:Pm_vs_fc}
\end{figure}

The stability of the above fixed points can be established by computing the eigenvalues of the stability matrix. After some tedious algebra, it can be shown that the above fixed points are stable in the three regions shown in Fig.~\ref{fig:Pm_vs_fc}(a) and defined by the following simple equations in the plane $(P_m,f)$:
\begin{align}
P_m&=P_m^c=1\\
f&=f_{c_1}=24\, \sqrt{2}\\
f&=f_{c_2}=12\, \sqrt{2}\, \frac{P_m+1}{P_m^{3/2}}
\end{align}
Since the velocity and magnetic field amplitudes $u_\alpha$ and $b_\alpha$ must be real numbers, the solutions \emph{Fluid A$^\pm$} only exist for $f>f_{c_1}$. They are stable for $P_m<1$. The solution \emph{Fluid B} is stable for $P_m<1$ and $f<f_{c_1}$,  and for $P_m>1$ and $f<f_{c_2}$. It is unstable elsewhere. The solution \emph{MHD$^\pm$} is stable for $P_m>1$ and $f>f_{c_2}$ and is unstable elsewhere. In summary, we have only fluid solutions for $P_m<1$, while an MHD solution is possible only for $P_m>1$ and $f>f_{c_2}$. The above solutions cover the entire $(f,P_m)$ parameter space. The system converges to one of these states depending on the parameter values irrespective of its initial conditions,  and the system is neither oscillatory nor chaotic. Clearly, the amplitude of $b_1$ and $b_2$ close to the dynamo threshold increases as $\sqrt{f-f_{c_2}}$. The dynamo transition in this low-dimensional model is thus a pitchfork bifurcation. 

The above results compare remarkably well with the numerical findings of Schekochihin et al.~\cite{Sche:dynamo_critPm_PRL04,Sche:DynamoApJ04} where a dynamo transition is also found for $P_m\ge P_m^{c}$ with $P_m^{c}$ near 1. Considering the simplicity of the above model, such a quantitative agreement may very well be fortuitous. Nevertheless, it is also interesting to notice that, in our model, $f_{c_2}$ decreases for increasing Prandtl numbers. Hence, it is easier to excite nonhelical dynamo for larger $P_m$ or more conductive magnetofluid. This feature of our model is also in agreement with recent numerical simulations~\cite{Sche:dynamo_critPm_PRL04,Sche:dynamo_lowPm_PRL07,Pont:dynamo_lowPr_PRL05,Mini:DynamoPoP06,Mini:LowPr} where an increase of the critical magnetic Reynolds number $Re_m^{c}$ is reported for smaller values of $P_m$.

It is also interesting to express the above results in terms of the Reynolds number instead of the forcing amplitude. A Reynolds number can be build by defining the large scale velocity as $U_L=\sqrt{u_1^2+u_2^2}\ u^\star$. Based on this velocity scale, the kinetic Reynolds number is given by:
\begin{equation}
Re=\frac{U_L}{\nu k_0}=\left\{
\begin{array}{ll}
\displaystyle{\frac{\sqrt{2}\, \sqrt{f^2-576}}{4}}& \emph{Fluid A}\\
\\
\displaystyle{\sqrt{2}\ \frac{s^2-12}{s}}& \emph{Fluid B}\\
\\
\displaystyle{\frac{f\ P_m}{2(1+P_m)}}& \emph{MHD}\\
\end{array}
\right.
\end{equation}
For large $f$, the amplitude of the magnetic field is proportional to the kinetic Reynolds number:
\begin{equation}
|b_1|\approx|u_1|\approx  Re /\sqrt{2}.
\end{equation}
This result is consistent with the result obtained by Monchaux et al.~\cite{Fauv:VKS_PRL06,Fauv:dynamo_reverse_EPL06,Fauv:Scaling07}. The critical Reynolds number for the dynamo transition between the \emph{Fluid B} and MHD solutions is easily computed:
\begin{equation}
Re^{c}=6\sqrt{\frac{2}{P_m}}
\end{equation}
and the critical magnetic Reynolds number $Re_m^{c}$:
\begin{equation}
Re_m^{c}=P_mRe=6\sqrt{2P_m}.
\end{equation}
Since $P_m>1$, $Re_m^{c}>6\sqrt{2}$. Note that $Re_m^{c}Re^{c}=72$. Hence, the $Re^{c}-Re_m^{c}$ curve is a hyperbola for $Re^{c}\le6\sqrt{2}$ since dynamo exists for $P_m>1$.

To gain further insights into the dynamo mechanism we have investigated the energy exchanges among the modes of the model. The fluid modes
$u_1$ and $u_2$ gain energy from the forcing $\mathbf{f}$ and give energy to the mode $u_3$. The magnetic modes $b_1$ and $b_2$ gain energy from the mode $u_3$. The net energy transfer to the mode $b_3$ through nonlinear interaction vanishes. Consequently the mode $b_3$ goes to zero due to dissipation. The $b_1$ and $b_2$ modes also lose energy due to dissipation. The energy balances for these modes reveal interesting feature of dynamo transition. At the steady-state, the energy input in the mode $b_1$ due to nonlinearity ($-b_1b_2u_3/\sqrt{6}=-b_1^2u_3/\sqrt{6}$) matches the Joule dissipation ($2b_1^{2}/P_m$) [Eq.~(\ref{eq:b101})]. For $f<f^{c2}$, the value of $u_3$ is less than $-2\sqrt{6}/P_m$, hence $b_1\rightarrow0$ asymptotically, thus shutting down the dynamo. For $f>f^{c2}$, $u_3=-2\sqrt{6}/P_m$, thus the energy input to the mode $b_1$ exactly matches with the Joule dissipation. This is the reason why $b_1$ and $b_2$ are constants asymptotically in the MHD regime.

We also studied a variant of the above model in which $f_1=0$ and $f_2 = f$.  For this model, the only solution is $u_2 = f/2$ with all the other variables being zero.  Hence this model does not exhibit dynamo. In the next subsection we will discuss another low-dimensional model that contains helicity.

\subsection{Helical model}

\begin{figure}[ht]
\begin{centering}
\includegraphics[width=0.8\columnwidth,keepaspectratio]{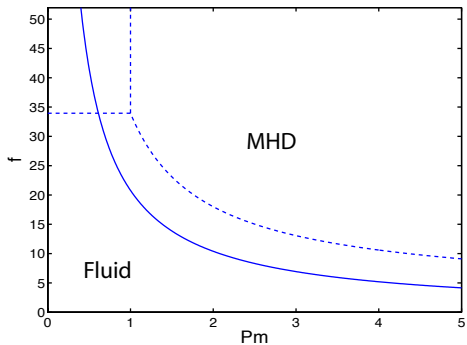}
\par\end{centering}
\caption{Plot of critical force amplitude $f_{c_3}$ as a function
of $P_m$ for the helical model. The dashed lines reproduce the stability regions of the non-helical model.}
\label{fig:Pm_vs_fcb}
\end{figure}

In this second version of the model, the value $h=1$ has been chosen. The helicity captured in the subspace ${\cal S}^<$ is then given by:
\begin{equation}
\langle \mathbf{u}^< \cdot (\nabla \times \mathbf{u}^<)\rangle=\left(\frac{4}{3}\ (u_1^2-u_2^2)-\frac{3}{2}\ u_3^2\right) \ (u^\star)^2\ k_0.
\label{eq:helicity_formula}
\end{equation}
If we restrict again the forcing to the large-scale modes ($f_3=0$), the amount of helicity carried on by the velocity field is expected to increase, at least in absolute value, if the forcing act differently on the modes $\mathbf{e}_1$ and $\mathbf{e}_2$. We have chosen the extreme case for which the forcing acts only on $u_2$ ($f_1=f_3=0$ and $f_2=f$):
\begin{equation}
\mathbf{f}=f\, \mathbf{e}_2
\end{equation}
where again $f=\sqrt{\langle \mathbf{f}\cdot\mathbf{f}\rangle}$. For these parameters, a unique fluid stationary solution and two stationary MHD solutions are found:
\begin{align}
\emph{Fluid}&
\left\{
\begin{array}{l}
u_1=u_3=0\\
\\
u_2=\displaystyle{\frac{f}{2}}\\
\end{array}
\right.
\label{eq:Solfluid2}\\
\nonumber\\
\emph{MHD$^\pm$}&
\left\{
\begin{array}{l}
u_1=u_3=0\\
\\
u_2=\displaystyle{\frac{6\ \sqrt{3}}{P_m}}\\
\\
b_1=\sqrt{3}\  b_3=\displaystyle{\pm \frac{\sqrt 3}{2}\ \left(-\frac{72}{P_m}+2\ \sqrt{3}\ f\right)^{1/2}}\\
\\
b_2=0
\end{array}
\right.
\label{eq:solMHD2}
\end{align}
All these solution obviously carry a non-zero resolved helicity as defined by~(\ref{eq:helicity_formula}). The fluid solution is stable for $f < f_{c_3}\equiv12\ \sqrt{3}\ /P_m$, while the MHD solutions are stable for $f > f_{c_3}$.  The plot of the critical force $f_{c_3}$ vs. $P_m$  for helical model is shown in Fig.~\ref{fig:Pm_vs_fcb}. Note that the helical model exhibits dynamo for all $P_m$ as long as $f > f_{c_3}$. 
Also, the critical forcing for the helical model is lower than the corresponding value for the nonhelical model. Hence it is easier to excite helical dynamo compared to the nonhelical dynamo consistent with recent numerical simulations \cite{Sche:dynamo_lowPm_PRL07,Pont:dynamo_lowPr_PRL05,Mini:DynamoPoP06,Mini:LowPr}.

Here again, the Reynolds number based on the large scale velocity $U_L =  u^*\, \sqrt{u_1^2+u_2^2}$ can be defined and yields:
\begin{equation}
Re=\frac{U_L}{\nu k_0}=\left\{
\begin{array}{ll}
\displaystyle{f/2} & \emph{Fluid}\\
\\
\displaystyle{\frac{6 \sqrt{3}}{P_m}} & \emph{MHD}\\
\end{array}
\right.
\end{equation}
and the magnetic Reynolds number for MHD case is $Re_m = 6 \sqrt{3}$, a constant.

Using Eq. (\ref{eq:helicity_formula}) we can compute the resolved kinetic helicity $H_K$ and current helicity $H_J$ defined as
\begin{eqnarray}
H_K  & = & \langle \mathbf{u}^< \cdot (\nabla \times \mathbf{u}^<)\rangle  =  
\frac{4}{3} u_2^2  (u^\star)^2\ k_0 \nonumber \\
& = &   -\frac{144}{P_m^2} (u^\star)^2\ k_0 \\
H_J & = & \langle \mathbf{b}^< \cdot (\nabla \times \mathbf{b}^<)\rangle  =  
\frac{5}{2} b_3^2  (u^\star)^2\ k_0 \nonumber \\
& = &  -\frac{5}{2}  \left( \frac{\sqrt{3}f}{2} - \frac{18}{P_m} \right)
(u^\star)^2\ k_0 
\end{eqnarray}
Pouquet et al.~\cite{Pouq:EDQNM}, and Chou \cite{Chou:theo} conjectured  the alpha parameter of dynamo to be of the form
\begin{equation}
\alpha \approx \alpha_u + \alpha_b = \frac{1}{3} \tau \left( -\mathbf{u} \cdot \nabla \times \mathbf{u} 
			+ \mathbf{b} \cdot \nabla \times \mathbf{b} \right),
\end{equation}
where $\tau$ is the velocity de-correlation time. Clearly $\alpha$ is optimal if $H_K<0$ and $H_J>0$.  Similar features are observed in flux calculation of Verma \cite{MKV:MHD_Helical}. These conditions are satisfied in the helical model indicating a certain internal consistency with the results of Pouquet et al.~\cite{Pouq:EDQNM}, and Chou \cite{Chou:theo}.

The energy exchange calculation of the helical model reveals that the magnetic modes $b_1$ and $b_3$ receive energy from the velocity mode $u_2$; the rate of energy transfer is proportional to $b_1 b_3 u_2$ that matches exactly with the Joule dissipation rate.  Since $u_2 = 6\sqrt{3}/P_m$ for all $P_m$, the rate of energy transfer $b_1 b_3 u_2$  matches with Joule dissipation rate ($\propto b_1^2/P_m$), and the dynamo is possible for all $P_m$ in case of helical model.  In contrast, in nonhelical model the corresponding velocity mode $u_3$ varies as $1/P_m$ for $P_m>1$, but saturates at $2\sqrt{3}$ for $P_m<1$, hence the energy transfer cannot match the Joule dissipation rate for $P_m<1$ for nonhelical model thus shutting off the dynamo for $P_m<1$.   This is one of the main difference in helical and nonhelical models.  Also note that as discussed at the end of the previous subsection,  the nonhelical model with the forcing $f_1=0, f_2=f$ does not exhibit dynamo for any parameter.

\section{Discussion}

In this paper, a class of low-dimensional models that exhibit dynamo transition is derived. These models depend on several parameters such as the Prandtl number, the forcing amplitude and a parameter, $h$, that characterizes the ability of the velocity modes to carry kinetic helicity. The fixed points of two simple models corresponding to $h=0$ and $h=1$ are studied in details. The first model is nonhelical (zero kinetic and current helicity) and is compatible with a stationary nonzero magnetic solution only for $P_m>1$. The second model has nonzero kinetic and current helicities, and it has nonzero stationary magnetic solution for all $P_m$. These findings confirm the idea that both kinetic and current helicities may play an important role in the dynamo transition, especially in helical model for $P_m<1$. These two values of $h$ correspond to the only values for which one of the nonlinear coupling in the low dimensional model vanishes. With the specific choice of the forcing proposed in the previous section, these models have then exact and tractable solutions. Arbitrary values of $h$ would lead to much more complex systems. 

Obviously, the models presented here are only a small subset of many possible MHD models that could exhibit dynamo. For instance, Rikitake~\cite{Rikitake:PhilSoc} constructed a dynamo model consisting of two-coupled disks that shows field reversal. Nozi\`eres \cite{Nozieres} proposed a model involving one velocity and two magnetic field variables. These two models are phenomenological. Recently,  Donner et al.~\cite{Donner:Lowdim_physd_2006} proposed a truncation of the MHD equations along the same lines as our model, but derived a much more complex system of equations for 152 modes.  Donner et al.~\cite{Donner:Lowdim_physd_2006} analyzed the dynamical evolution of these modes for $P_m=1$ and observed steady-state and chaos in their system. The main advantage of our model is that it allows a complete analytical treatment. 

The six variables of our model are only representative of the large-scale modes of the system, while a realistic description of a turbulent system exhibiting dynamo should have a large number of modes. Although the small scale variables are often quite important in turbulent flow, in some cases the large scale variables may determine some of the flow properties. The surprising success of the very simple models proposed here in reproducing several feature of the dynamo transition could suggest that we are in such a situation. Schekochihin et al.~\cite{Sche:dynamo_critPm_PRL04,Sche:DynamoApJ04}, Stepanov and Plunian \cite{Step:DynamoGrowth}, and Iskakov et al.~\cite{Sche:dynamo_lowPm_PRL07} have highlighted the role played by inertial range eddies that are absent in our low-dimensional models.  Yet in the absence of a definite theory of dynamo, it is interesting to show that low-dimensional models that focus on the dynamics of the large scale flows may also be successful.

\section*{Acknowledgements}

This work has been supported in part by the Communaut\'e Fran\c caise de Belgique (ARC 02/07-283) and by the contract of association EURATOM - Belgian state. The content of the publication is the sole responsibility of the authors and it does not necessarily represent the views of the Commission or its services. D.C. and T.L. are supported by the Fonds de la Recherche Scientifique (Belgium).  MKV thanks the Physique Statistique et Plasmas group at the University Libre du Brussels for the kind hospitality and financial support during his long leave when this work was undertaken.  This work, conducted as part of the award (Modelling and simulation of turbulent conductive flows in the limit of low magnetic Reynolds number) made under the European Heads of Research Councils and European Science Foundation EURYI (European Young Investigator) Awards scheme, was supported by funds from the Participating Organisations of EURYI and the EC Sixth Framework Programme.

\end{document}